# GREEN COMPUTING IN DEVELOPED AND DEVELOPING COUNTRIES

Dr.S.Taruna[1], Pratibha Singh[2] and Soshya Joshi[3]

[1]Associate Professor, Banasthali Vidyapith, Rajasthan, India
[2,3]M.Tech Student, Banasthali Vidyapith, Rajasthan, India

*ABSTRACT*

*Today e-waste are becoming a major problem for the developing countries. E-waste is defined something as a discarded parts of electronic devices which contains most of the times, hazardous chemicals which is deadly for our environment, example is computer components. Green Computing is the study and practice of designing, using, disposing and manufacturing electronic components in an eco-friendly manner and Green Computing is one of the solution to tackle with this hazardous e-waste problem which is an emerging concern towards the environment. The objective of this paper is to draw the attention towards the lack of awareness about green computing or we can say how green computing policies is being ignored by developing countries and how developed countries are adopting green IT policies seriously. This paper also discusses the analysis which has been done on how the amount of e-waste has been increased in developing countries in past years.*

## I. INTRODUCTION

Today the requirement of new electronic devices are growing very rapidly day by day and even the production of these electrical and electronic devices is growing very rapidly, to meet the requirements of consumers worldwide. As, every day the new upgraded devices are replacing the old ones and the devices which has been replaced are dumped over somewhere and not been taken in use in any manner. Even though to environment and to human being this is a serious problem but some of developed nations have already been taken steps towards this type of serious concern but the question arises are the less fortunate developed nations actually taking any concern or steps to totally eradicate e-waste problem from their land? Recently climate scientist have proposed some results which is quite shocking- percentage of greenhouse gases (GHG) has so much increased that it has crossed what was originally predicted. Target of scientists, economists and policy makers is to reduce emissions 20% below 1990 levels in 2020 [6].

## II. GREEN COMPUTING

Green Computing is an efficient way toward energy efficient products .Green computing is an environmentally responsible use of computers and related resource. The aim of green computing is to lessen unsafe, unhealthy material, increase energy efficacy during products lifetime and boost the recyclability of products [1]. If we seriously want to ponder over environmental impacts-then our foremost focus should be on manufacture, which has the greatest impact over environment. Main focus should be given to manufacture to correct this condition , as 80% of energy emission is during manufacture only, that means which is much before a PC is turned on.

                                                                                                              97

Waste is only one part of a product life cycle, tackling the waste problem requires looking beyond simply end of pipe solutions. Programmes like recycling, which is very important but that doesn't means will alone minimize the production of waste. In design phase only we must make major decisions- like which substances should be used, whether usage of that substance is risky or not, what is the products longevity and recyclability and overall what impact this product has on environment[15].

## III. PRESENT SCENARIO

These days with rapid increase in technology, electronic products are used for short span and older products are dumped frequently. One reason is replacement is much cheaper than repairing of electronic gadgets. According to US environmental protection agency many electronic goods which are not in used i.e. about $3/4^{th}$ in amount were thrown away either dumped in waste, unused, scrappy land i.e. Landfills or destroy by burning or exported to different countries such as Asian countries.

If dumped in unused lands then the toxic chemicals produced by these electronic product either release in the atmosphere or get mix into soil and water. In European countries measures have been taken to prevent such type of electronic waste disposing .By burning chemicals like lead, mercury, cadmium-ashes are mixing in air which is so much harmful that it ill effect food chain. As, while burning PVC plastic which is present in making computer and in many electronics components produces dioxins and furame which again is hazardous.

Another method followed by developing countries these days is to export their e-waste to developing countries. Recycling which is taking place over this electronic waste is a good thing, but more that the disappointing point is the way the raw materials from these products are processed and handled, which can harm the workers who are handling these products while recycling as well as is harmful to environment.

The difference between the recycling process in developed and developing countries is ...that, in developed countries like Europeans parts, some of the e-waste like plastics are not reprocessed to evade brominated furans and dioxins being released. While in developing countries there is no such steps taken over recycling. In a survey of 2005, 18 Europeans seaports were inspected and it is found that about 475 of e-waste was tried to export illegally. This is also a major concern that violation of international law are also taking place to get rid of e-waste to such an extent that in UK alone at least 23,000 metric tons was illegally shipped to India , Africa and China.[14]

In this way, in US also it is found that 50-80% of the waste should be used for recycling in their country itself is illegally exported, China tried to solved this major problem by banning the import of e-waste, but these law are not working as still tones of e-waste is imported illegally. A best example is GUIYA of GWANGDONG province, India also facing this e-waste trade problem. Many scrap yards can be seen in Delhi, Mumbai, Bangalore, Firozabad, Meerut, were thousands of workers are leading perished life because they are thinking to earn for today leaving behind the thought that what will happen to them tomorrow.

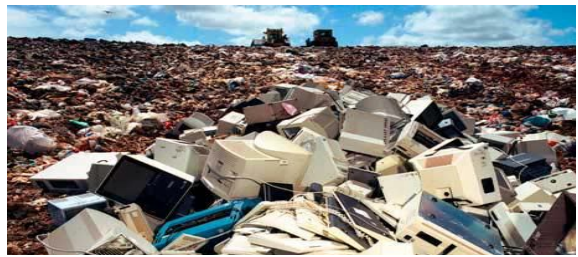

Fig.1. Landfills due improper disposal of computers.

If they continue to live in between hazardous chemicals. Now, why developed countries are dumping, exporting there e-waste in developing countries, answer is the low cost of recycling.
To recycle a glass of monitor in US it will cost 10 times much then in any developing countries, and why developing countries are importing these e-waste answer is –recycling of this e-waste can extract substance like copper, nickel, iron, gold, silicon-by selling these metals scraps recyclers can make money.

It is cheaper to have the hard labour of pulling apart and melting down pieces done put side the country even if that means the useless scraps and other hazardous materials will litter that area.[3] And in developing countries unlike developed countries laws, are also inadicate. People are taking job in there scrap yards because it gives them decent wages to live and often children are taking up their jobs, which are very dangerous for their health. Environment in developed countries is paying more focus on priority over the profit drawing specially in industry and economy, and innovation in own native land than enforcement of laws.

It is not better to ship developed countries electronic waste to other countries for disposal –rather than-proper methods should be made to recycle maximum components  used in manufacturing and usage of minimum hazardous chemicals and last even if to dispose there e-waste , disposal should be complete 100% ecofriendly way.

It will be better to use different chemical composition for new discovery so that single chemical won't reside in atmosphere in abundance, because the abundance presence of particular element can cause major dangerous effects.

One way which European union has adopted, which is quite impressive also is- after manufacturing an equipment, manufacturer is responsible financially or physically for their equipment, till their survivability- which arises a competitive encouragement for companies to plan greener products [3].

The methods, operations which China, India and Pakistan has adopted- that is the e-w2aste recycling and discarding are extremely polluting and is terribly unhealthy for humans. Examples including open burning of plastic waste, exposure of toxic solders, river dumping of acids, and widespread general dumping.

In developing countries stakeholders customers are looking for good facility but cheaper product. According to Basel Action Network executive director Jim Pockett-  when an electronic waste is taken for recycler, 80% of that material is loaded in container ship, going to countries like India, China, Pakistan, where the worst happens to it [3].

In developing countries stakeholders /customers are looking for heaper good facility profitable product , when the cheap word comes means comprising with the elements used in the manufacture of product , as star products are little costly but more environment friendly ;this is also a great problem as in developing countries versus developed countries value of currencies has a huge difference which also plays a key role in choosing the product , as well as makes a great gap in the private sector businesses.

## IV. POLICIES IN DEVELOPED COUNTRIES

Many government programs around the world today focused on environmental sustainability, are exploring technology initiatives for reducing greenhouse gases. For example the ministry of science, technology and innovation of Denmark's efforts for green IT establishment and another example is from the ministry of economy, trade and industry of Japan- which established green IT initiative, which provide a strong model of green innovation policy.

The two organizations of US example US Department of Energy (DoE) and the US Environmental Protection Agency (EPA) has also initiated in the field of green ICT. Technology may be important to green innovation but strategic public policy is critical. To encourage clean energy project the US administrations has dedicated US $71 billion to clean energy funding with a supplementary of US $20 billion for loan assurance and tax stimulus [16].

The climate change at (2008) sets a legally binding objective for diminish UK carbon dioxide emissions by 80% from 1990 levels by 2050[17].

In Australian Government's ICT sustainability plan 2010-15 Tony Chan pointed plans and actions for the agencies to lessen their release [18].

The RoHS (Restriction of the use of certain hazardous Substances in Electrical and Electronic equipment) laws (directive 2004/95/EC), in force since 13 February 2003, is marked as the first law in the world that limits the use of hazardous substances in electrical and electronic equipment. Six toxic substances have been constricted: lead, mercury, cadmium, hexavalent chromium, polybrominated biphenyls (PVV) and poly brominated diphenyl ethers (PEDS). For cadmium the maximum value which is been set is 0.01%, which is an exception case and for others it is been considered that value should not exceed 0.1% .

The WEEE (Waste Electrical and electronic equipment) legislation (directive 2002/96/EC) has also come in force on 13 February 2003.This regulation is about collection, recycling and regaining of electronic goods. It is based on take back system where the used products can be returned by the consumers which is free of cost and the authority is given to the production team for managing e-waste properly[4].

Green grid consortium is a tremendous initiative. It is a global association of companies devoted to developing and endorse standards, measurement method, processes and new technologies that lead to energy efficiency in data centers [19].

## V. POLICIES IN DEVELOPING COUNTRIES

In Copenhagen summit, India has committed to minimize IT emission by 20-25% as compared to the 2005 emission levels. For Green India, Government of India, ministry of environmrnt and forest(2011) has also initiated few steps under National Action Plan on Climate Change(NAPCC). They have proposed tentative action plan for implementation of Green India mission during 2011 to 2012[20].

To control an increasing problem in Africa, Safaricom has initiated an e-waste scheme in Kenya, together with South Africa. In Africa most of the recycling is done on an easy go, casual, unconstrained basis often in wild ungoverned dumpsites or landfills. The issue is that most African countries do not yet have policies in place to support the established of e-waste plants. Zambia alone has about 10 million mobile users, the estimation is given by the Zambia Information and Communication Technology Authority(ZICTA) which is the regulator of countries telecom sector. Most of the countries are counterfeit devices from China that just last a few months and are disposal of carelessly. AS compared to other regions the usage of electric and electronic equipment is not to that extent high but still it is increasing in an unpleasant phase [21].

## VI. SOLUTIONS TO BE ADOPTED

As e-waste are originating day by day in developed countries and even becoming problems for everyone, so now it's time to reduce the impact of e-waste and to take a important steps towards Green Computing to make our environment clean and free from all this kind of toxic chemicals.
One of the solutions regarding this be as such the countries especially focusing towards the developing countries must have to go through the policies strictly or to adopt these policy seriously which the government have been promised. The developing countries must also avoid into take a waste from the developed countries at low prices and then using them for manufacturing of some other products which cost them cheap as doing so.

Another solution can be with the contribution of citizens as well as the manufacturers/producers. The citizens or consumers can also contribute towards this by making their preferable choice to buy those electronic products which can be recycled or it can be reuse and also those which contains a less toxic chemicals.

In educational institutes make Green IT subject a compulsory one rather than an optional one, so that new ideas can be developed by students , based on Green IT and with that innovation in this field that could make products more cheaper and less hazardous and should have the abilities to attract more customers than normal products. By introducing in educational institutions is the only way to invite more project and ideas.

Awards, scholarships should be made to increase more and more contribution in this field. Awareness among public is required but what if cheaper healthy products are launched in market definitely people will get attracted to buy them.

Like on every health hazardous products warning is return same way on electronic gadgets it should be written weather it is green product or not.

Government should charge extra tax on those companies which are not following Green ICT rules as well as not producing Green ICT products.

## VII. CONCLUSION

From this paper we can conclude that the developing countries are saying that they are opting green IT policies but in actual things are going in reverse direction. Developing countries need to seriously think about their policies and strategies. More and more attention is required in this field and our main aim is to showcase the present scenario.

## Authors

Dr. S.Taruna is an active researcher in the field of communication and mobile network, currently working as Associate Professor in Department of Computer Science at Banasthali University (Rajasthan), India. She has done M.Sc from Rajasthan University and PhD in Computer Science from Banasthali University (Rajasthan), India. She has presented many papers in National and International Conferences, published around 25 papers in various journals and reviewer of various journals and conference.

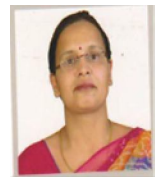

Pratibha Singh is pursuing M.Tech in Information Technology from Banasthali University, and done B.Tech from V.B.S Purvanchal University in 2013.

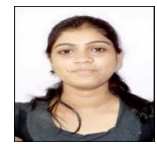

Soshya Joshi is pursuing M.Tech inInformationTechnologyfrom Banasthali University, and done BE from Rajiv Gandhi Technical University, Bhopal.

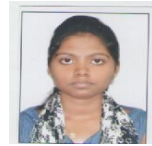